\documentclass[camera,letterpaper,nomarginnotes,nonarrowgutter]{jpaper}

\usepackage{dblfloatfix}
\usepackage{multirow}
\usepackage{booktabs}
\usepackage{array}
\usepackage{cite}
\usepackage{algorithm}
\usepackage[noend]{algpseudocode}
\usepackage[rightComments=true,commentColor=gray]{algpseudocodex}
\usepackage{eucal}
\usepackage{rotating}
\usepackage{xcolor,colortbl}
\definecolor{freakishgreen}{HTML}{0A982B}
\definecolor{urlblue}{HTML}{319dd6}
\usepackage{ifthen} %
\usepackage{xspace} %
\usepackage{enumitem}%
\usepackage{fancyhdr}
\usepackage{enumitem}%
\usepackage[noend]{algpseudocode}
\usepackage{datetime}
\usepackage{textcomp} %
\usepackage{xparse} %
\usepackage{setspace} %
\usepackage{listings} %
\newcolumntype{L}[1]{>{\raggedright\let\newline\\\arraybackslash\hspace{0pt}}m{#1}}
\newcolumntype{C}[1]{>{\centering\let\newline\\\arraybackslash\hspace{0pt}}m{#1}}
\newcolumntype{R}[1]{>{\raggedleft\let\newline\\\arraybackslash\hspace{0pt}}m{#1}}
\newcolumntype{M}[1]{>{\centering\arraybackslash}m{#1}}

\usepackage{tikz}
\newcommand*\circled[1]{\tikz[baseline=(char.base)]{
            \node[shape=circle,fill,inner sep=1pt] (char) {\textcolor{white}{#1}};}}

\makeatletter
\def\thickhline{%
  \noalign{\ifnum0=`}\fi\hrule \@height \thickarrayrulewidth \futurelet
   \reserved@a\@xthickhline}
\def\@xthickhline{\ifx\reserved@a\thickhline
               \vskip\doublerulesep
               \vskip-\thickarrayrulewidth
             \fi
      \ifnum0=`{\fi}}
\makeatother
\newlength{\thickarrayrulewidth}
\usepackage{pifont}

\def\thickhline{\noalign{\hrule height.8pt}}
\newcolumntype{?}{!{\vrule width 0.8pt}}

\newcommand{\paraheading}[1]{\vspace{2pt}\noindent \textbf{#1}}
\newcommand{\paraspace}{\vspace{2pt}\noindent}

\usepackage{xhfill}

\usepackage{soul}
\setul{0.5ex}{0.3ex}
\setulcolor{blue}

\newif\ifsubmission
\submissiontrue

\newcommand{\rbc}[1]{{#1}}
\newcommand{\edit}[1]{{#1}}
\newcommand{\rbone}[1]{\textcolor{BurntOrange}{#1}}

\newcommand{\rn}[1]{\textcolor{violet}{#1}}

\newenvironment{revblock}
{\ifsubmission\ignorespaces\else\color{Blue}\ignorespaces\fi}
{\ifsubmission\ignorespaces\else\normalcolor\ignorespaces\fi}

\definecolor{colorV1}{rgb}{0.0, 0.0, 1.0}    %
\definecolor{colorV2}{rgb}{1.0, 0.0, 1.0}    %
\definecolor{colorV3}{rgb}{0.0, 0.5, 0.0}    %
\definecolor{colorV4}{rgb}{0.6, 0.4, 0.2}    %

\newcounter{CurrentDraftVersion}

\ExplSyntaxOn

\cs_new:Npn \revcol:n #1
  {
    \int_case:nnF {#1}
      {
        {1}{blue}
        {2}{magenta}
        {3}{orange}
        {4}{cyan}
        {5}{violet}
        {6}{teal}
        {7}{purple}
        {8}{brown}
        {9}{olive}
        {10}{red}
      }
      {blue}
  }

\NewDocumentCommand{\chg}{ O{\value{CurrentDraftVersion}} m }
  {
    \int_compare:nNnTF { \value{CurrentDraftVersion} } = {1000}
      { #2 }
      {
        \int_compare:nNnTF { \value{CurrentDraftVersion} } = {100}
          { \textcolor{blue}{#2} }
          {
            \int_compare:nNnTF { #1 } = { \int_eval:n { \value{CurrentDraftVersion} - 1 } }
              { \textcolor{\revcol:n {#1}}{#2} }
              { #2 }
          }
      }
  }

\ExplSyntaxOff

\ifsubmission
    \renewcommand{\rbone}[1]{{#1}}
    \renewcommand{\rn}[1]{{#1}}
    
    \fancypagestyle{firstpage}
    {
        
        \fancyhf{}
        \fancyfoot{}
        \fancyhead[C]{This is an extended and updated version of an article published in 
        \\ IEEE Micro 2026 \url{https://doi.org/10.1109/MM.2026.3667076}}
        \fancyfoot[C]{\thepage}
    }
    
\else
    
    \fancypagestyle{firstpage}
    {
        \fancyhead[C]{\textcolor{blue}{\emph{Version \arabic{CurrentDraftVersion}~---~\today, \xxivtime \ UTC}}}
        \fancyfoot[C]{\thepage}
    }
\fi

\usepackage{hyperref}

\hypersetup{
    bookmarks=true,
    breaklinks=true,
    colorlinks=true,
    linkcolor=blue,
    citecolor=blue,
    urlcolor=urlblue,
    pdfpagelabels=false,
}

\usepackage[nameinlink]{cleveref}
\crefformat{section}{\S#2#1#3} %
\crefformat{subsection}{\S#2#1#3}
\crefformat{subsubsection}{\S#2#1#3}

\lstdefinestyle{shell}{
  basicstyle=\ttfamily\small,
  columns=fullflexible,
  keepspaces=true,
  breaklines=true,
  showstringspaces=false,
  upquote=true,
  frame=none,        %
  xleftmargin=0pt,   %
  aboveskip=6pt,
  belowskip=6pt
}

\def\BibTeX{{\rm B\kern-.05em{\sc i\kern-.025em b}\kern-.08em
    T\kern-.1667em\lower.7ex\hbox{E}\kern-.125emX}}

\title{\LARGE{Machine Learning-Driven Intelligent Memory System Design: From On-Chip Caches to Storage}}

\author{
    Rahul Bera \hspace{0.5em} Rakesh Nadig \hspace{0.5em} Onur Mutlu\vspace{0.5em} \\
    \normalsize{
        SAFARI Research Group, ETH Zürich, Switzerland
    }
}

\begin{document}

\maketitle
\thispagestyle{firstpage}

\begin{abstract}
Despite the data-rich environment in which memory systems of modern computing platforms operate, many state-of-the-art architectural policies employed in \chg[1]{the} memory system rely on static, human-designed heuristics that fail to \chg[1]{truly} adapt to the workload and system behavior \chg[1]{via principled learning methodologies}. 
In this article, we propose a fundamentally different design approach: using lightweight and practical machine learning (ML) methods to enable adaptive, data-driven control throughout the memory hierarchy.

We present three ML-guided architectural policies: (1) \emph{Pythia}, a reinforcement learning-based data prefetcher for on-chip caches, (2) \emph{Hermes}, a perceptron learning-based off-chip predictor for multi-level cache hierarchies, and (3) \emph{Sibyl}, a reinforcement learning-based data placement policy for hybrid storage \chg[1]{systems}. 

Our evaluation shows that Pythia, Hermes, and Sibyl significantly outperform the best-prior human-designed policies, while incurring modest hardware overheads. 

Collectively, this article demonstrates that integrating adaptive learning into memory subsystems can lead to intelligent, self-optimizing architectures that unlock performance and efficiency gains beyond what is possible with traditional human-designed approaches.
\end{abstract}

\section {INTRODUCTION}

{T}oday's computation is heavily bottlenecked by data. 
Many key applications (e.g., machine learning, graph analytics, large-scale recommendation systems, genome sequencing and analysis), irrespective of whether they are executed in cloud servers or mobile devices, are all data-intensive~\cite{oliveira2022accelerating,singh2021fpga,alser2020accelerating,amiraliphd,boroumand2018google,boroumand2021google,mutlu2020intelligent,mutlu2025memory,mutlu2022modern,mutlu2021intelligent,boroumand2022polynesia, boroumand2019conda, he2025papi, Gu2025PIMIA, ghiasi2022genstore, ghiasi2024megis, ghiasi2026sage, mao2022genpip, alser2022molecules, soysal2025mars, mutlu2013memory,bera2025mitigating}.
They operate on large amounts of application data that overwhelm the storage and retrieval capability of high-performance memory systems employed in today's processors, rendering \chg[1]{memory} the key performance and energy bottleneck~\cite{mutlu2020intelligent,mutlu2021intelligent}.
To alleviate this, architects have progressively optimized the memory system with numerous sophisticated speculative (e.g., prefetching data to on-chip caches before the processor demands it~\cite{jouppi_prefetch}) and control (e.g., placing a data block in an appropriate device in \chg[1]{a} hybrid memory/storage system~\cite{singh2022sibyl,rakesh2025harmonia}) policies to reduce the data access \chg[1]{overheads}.

Even though these architectural policies observe a vast amount of application data (as well as system-generated metadata) during their online operation, we observe that these policies often make decisions based on simple, rigid \chg[1]{heuristics}, \chg[1]{unable to learn in a principled manner from} massive amounts of easily-available data. 
This is primarily because most existing architectural policies make \emph{human-design-driven} decisions, which predominantly rely on fixed and often myopic human-crafted heuristics \edit{that provide limited adaptability to complex system state and workload demands}. 
For example, architects have proposed numerous data prefetching policies to predict the addresses of future memory requests and fetch their data into traditional on-chip caches before the processor demands it.
Yet, most prefetching policies rely on a single piece of program context information (e.g., the program counter value of a load instruction), that a human architect selects at design time, to \chg[1]{identify} patterns in a program’s memory access stream.
As a result, these prefetching policies often fail to adapt to changing workload behavior across a diverse set of workloads~\cite{pythia}.

\textbf{Our goal} is to improve upon many such rigid human-designed policies found in modern state-of-the-art memory system designs by making them \emph{data-driven} and \chg[1]{fundamentally} adaptive \chg[1]{in a principled manner}. 

\chg[1]{To this end}, we propose harnessing lightweight and practical machine learning (ML) to guide architectural decision-making, enabling intelligent, data-driven control policies throughout the memory hierarchy: from the on-chip caches to the storage subsystem.
\chg[1]{Specifically}, we \chg[1]{describe} three ML-driven architectural policies: (1) \emph{Pythia}~\cite{pythia}, a reinforcement learning (RL)-based framework for intelligent hardware prefetching in on-chip caches, (2) \emph{Hermes}~\cite{hermes}, a perceptron learning-based mechanism for accelerating long-latency memory requests via accurate off-chip load prediction in a multi-level cache hierarchy, and (3) \emph{Sibyl}~\cite{singh2022sibyl}, \chg[1]{an} RL-based adaptive data placement policy for hybrid storage systems.

Our evaluation shows that Pythia, Hermes, and Sibyl not only provide significant performance \chg[1]{and energy improvements} on top of a well-optimized memory hierarchy design, but also consistently outperform prior best human-designed policies across a wide range of workloads and system configurations - owing to their adaptive \chg[1]{and principled} data-driven decision-making. 
With detailed hardware design and prototyping, we also show that all three proposed policies not only incur modest area and power overheads, but can be practically employed in today's computing systems.

Collectively, these proposals illustrate the significant potential of ML-driven policies for designing adaptive and self-optimizing memory systems, fundamentally transforming architectural decision-making processes.

\paraspace We make the following key contributions in this article:
\begin{itemize}
\item We analyze various human-designed state-of-the-art architectural policies proposed for three \chg[1]{components in} the memory system: data prefetching, off-chip prediction, and data placement. We show that these policies leave a significant performance potential on the table due to \chg[1]{their} rigid and myopic \chg[1]{designs}.
	
\item We propose three novel policies, Pythia, Hermes, and Sibyl, that employ various forms of machine learning (reinforcement and perceptron learning) to make adaptive \chg[1]{and principled} data-driven decisions online.

\item We show that Pythia, Hermes, and Sibyl provide significant \chg[1]{benefits} over the best human-designed policies across a wide range of workloads and system configurations.

\end{itemize}

\section{SHORTCOMINGS OF HUMAN-DESIGNED MICROARCHITECTURAL POLICIES}

\begin{figure*}[!ht]
    \centering
    \includegraphics[width=\textwidth]{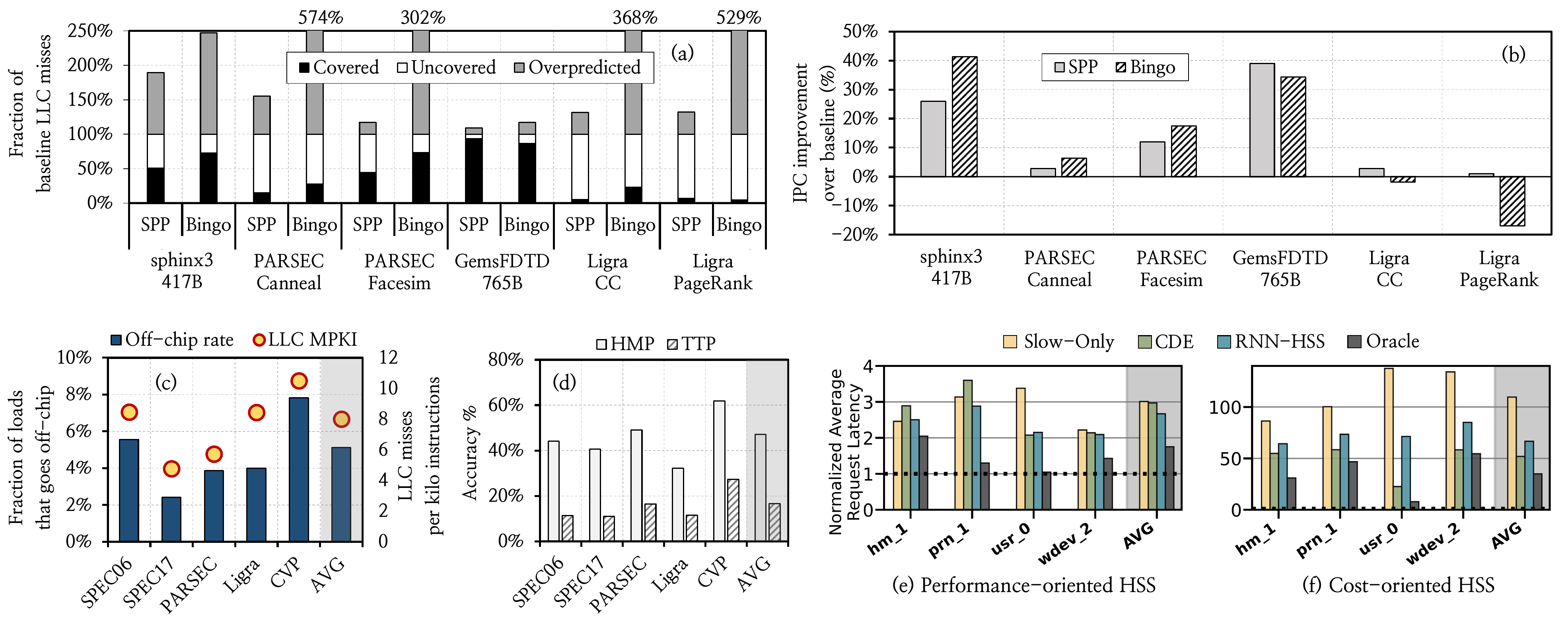}
    \caption{(a) Coverage, overprediction, and (b)
performance comparison of two recently-proposed prefetchers: SPP and Bingo. (c) Percentage of loads that miss the LLC and go off-chip (on the left y-axis) and the LLC MPKI (on the right y-axis)
in the baseline system with a state-of-the-art prefetcher. \edit{(d)} Accuracy of two off-chip predictors, HMP and TTP. Average request latency of CDE and RNN-HSS on (e) performance-oriented and (f) cost-oriented HSS. The average request latency is normalized to Fast-Only policy. \chg[1]{Figures adapted from our MICRO 2021~\cite{pythia}, MICRO 2022~\cite{hermes}, and ISCA 2022~\cite{singh2022sibyl} papers.}}
    \label{fig:motivation}
\end{figure*}

The ever-growing data footprint of many \chg[1]{modern} workloads (and likely future workloads) greatly overwhelms the storage and retrieval capability of the memory systems of modern machines~\cite{warehouse,cloudsuite,wang2014bigdatabench,boroumand2018google,boroumand2021google,alser2020accelerating,cali2020genasm}.
As such, data \chg[1]{access has become} the key performance and energy bottleneck~\cite{mutlu2013memory,mutlu2025memory,mutlu2021intelligent,mutlu2022modern}.
In response, architects have developed numerous speculative (e.g., data prefetching) and control (e.g., data placement and management) policies to mitigate data access \chg[1]{overheads}. 
Although these policies observe a vast quantity of application data and system-generated metadata during their online operation, their decision-making process often relies on simple and often myopic human-designed heuristics, failing to dynamically adapt to complex system states and evolving workload demands \chg[1]{in a principled manner}~\cite{rlmc,morse,mutlu2021intelligent,bera2025mitigating}. As a result, they often leave a large potential for performance and energy efficiency improvement on the table~\cite{mutlu2021intelligent,bera2025mitigating}.

In this section, we motivate the need for \emph{data-driven} policies by examining the shortcomings of conventional human-design-driven policies through three case studies that span the memory hierarchy: from the on-chip caches to the storage system.

\subsection{Case Study 1: Prefetching in On-Chip Caches}

Prefetching is a well-studied speculation technique that predicts the addresses of long-latency memory requests and fetches their corresponding data from the main memory to on-chip caches before the processor demands it~\cite{stride,streamer,baer2,jouppi_prefetch,ampm,fdp,footprint,sms,spp,vldp,sandbox,bop,dol,dspatch,bingo,mlop,ppf,ipcp,jimenez_cache_pref1,jimenez_pref1,jimenez_pref3,jimenez_pref4,Zhang2025HierarchicalPA,kpc,cooksey2002stateless,ebrahimi2009techniques,charney1995,charneyphd,iacobovici2004effective,panda2012prefetchers,panda2014xstream,markov,stems,somogyi_stems,wenisch2010making,domino,isb,misb,triage,wenisch2005temporal,chilimbi2002dynamic,chou2007low,ferdman2007last,hu2003tcp,bekerman1999correlated,karlsson2000prefetching,varkey2017rctp}.
To identify patterns within a program’s memory access stream, a prefetcher typically correlates memory addresses with program context information (also called \emph{program feature}).

Numerous prefetching techniques proposed in the literature consistently exhibit two primary limitations:
(1) reliance on a \emph{single, static, human-designed} program feature to identify memory access patterns, significantly limiting their efficacy across diverse workloads~\cite{stride,streamer,baer2,stride_vector,jouppi_prefetch,ampm,fdp,footprint,sms,sms_mod,spp,vldp,sandbox,bop,dol,dspatch,mlop,ppf,ipcp,navarro2022berti,chen2025gaze,pmp}, and
(2) lack of \emph{inherent system awareness} (e.g., memory bandwidth usage), resulting in performance degradation in resource-constrained systems~\cite{fdp,ebrahimi2009coordinated,ebrahimi2009techniques,ebrahimi_paware,dspatch,lee2011prefetch,lee2008prefetch,mutlu2005,wflin,zhuang,wflin2,charney,memory_bank,Lee2010DRAMAwareLC2,Lee2010DRAMAwareLC}.

To illustrate this, we show the coverage (i.e., the fraction of
\chg[1]{program} memory requests \chg[1]{correctly} predicted by the prefetcher) and overpredictions (i.e., prefetched memory requests that are not demanded by the \chg[1]{program}) of two
recently proposed prefetchers, signature path prefetcher (SPP~\cite{spp}) and Bingo~\cite{bingo} for six example workloads in~\Cref{fig:motivation}(a) and their performance improvement in~\Cref{fig:motivation}(b). 
We make two key observations from this figure.
First, since SPP and Bingo rely on two different yet static program features, \chg[1]{neither} of them provides the best performance benefit even across this small set of workloads.
Second, even if Bingo achieves a similar prefetch coverage in \texttt{Ligra-CC} as compared to \texttt{PARSEC-Canneal} while generating significantly lower overpredictions,
Bingo \emph{degrades} performance by $1.9$\% in \texttt{Ligra-CC} but \emph{improves} performance by $6.4$\% in \texttt{PARSEC-Canneal}, over a no-prefetching baseline.
This contrasting outcome is due to Bingo's lack of awareness of memory bandwidth usage \chg[1]{of the system}.

We conclude that, despite extensive research in prefetching, the static and human-designed nature of most conventional prefetching policies significantly limits their efficacy over a wide range of workloads and system configurations.

\subsection{Case Study 2: Off-Chip Prediction in Multi-Level Cache Hierarchies}

To cater \chg[1]{to} the ever-increasing data \chg[1]{footprints} of modern workloads, architects continue to increase the size (and also number of levels) of the on-chip caches in commercial processors, which inadvertently increases on-chip cache access \chg[1]{latencies}.
As a result, a large fraction of the latency of an off-chip memory request is spent accessing the on-chip cache hierarchy to solely determine that it needs to go off-chip~\cite{l3_lat_compare1,llc_lat3,hermes,d2d,d2m,lp,mnm}.

To mitigate this, architects have proposed \emph{off-chip prediction}, a technique that predicts which memory requests might go off-chip and speculatively fetches its data \emph{directly} from the main memory~\cite{hermes}.
However, predicting which memory requests will miss the entire cache hierarchy is challenging, especially due to the presence of complementary speculative techniques like data prefetching.
Our analysis, as depicted in~\Cref{fig:motivation}(c), shows that on average only $5.1$\% of the total loads generated by a workload
go off-chip in presence of a sophisticated data prefetcher.
\chg[1]{While researchers} have proposed techniques to accurately predict off-chip memory requests~\cite{yoaz1999speculation,lp,alloy_cache,d2d,d2m}, these techniques either rely on a fixed set of human-crafted program features to correlate with the off-chip requests~\cite{yoaz1999speculation} or expensive hardware structures to track cacheline tags in the entire cache hierarchy~\cite{lp,alloy_cache,d2d,d2m}. Yet, they fall short in providing accurate off-chip prediction across a wide range of workloads.
As~\Cref{fig:motivation}(d) shows, the prior-best \edit{hit-miss predictor} (HMP~\cite{yoaz1999speculation}) and the \edit{tag-tracking-based predictor} (TTP~\cite{lp,hermes}) provide only \chg[1]{$47$\%} and $16.6$\% accuracy \chg[1]{(i.e. the fraction of predicted off-chip loads that actually go off-chip)}, respectively.

This shows that the current off-chip predictors have a significant room for improvement due to their reliance on fixed, human-crafted program features.

\subsection{Case Study 3: Data Placement in Hybrid Storage Systems}
This case study explores the storage system, another critical component of the memory hierarchy. Modern high-performance systems employ hybrid storage, combining fast-yet-small storage devices (e.g., SSDs) with large-yet-slow devices (e.g., HDDs) to deliver high capacity at low latency~\cite{
meza2013case,bailey2013exploring,smullen2010accelerating,lu2012pram,tarihi2015hybrid,xiao2016hs,wang2017larger,lu2016design,luo2015design,srinivasan2010flashcache,reinsel2013breaking,lee2014mining,felter2011reliability,bu2012optimization,canim2010ssd,bisson2007reducing,saxena2012flashtier,krish2016efficient,zhao2016towards,lin2011hot,chen2015duplication,niu2018hybrid, oh2015enabling,liu2013molar,tai2015sla,huang2016improving,kgil2006flashcache,kgil2008improving,oh2012caching,yang2013hec,ou2014edm,appuswamy2013cache,cheng2015amc,chai2015wec,dai2015etd,ye2015regional,chang2015profit,saxena2014design,li2014enabling,zong2014faststor,do2011turbocharging,lee2015effective,baek2016fully,liu2010raf,liang2016elastic,yadgar2011management,zhang2012multi,klonatos2011azor}. 
The key challenge in designing a high-performance, cost-effective hybrid storage system (HSS) is to accurately identify the performance-critical application data and place it in the best-fit storage device~\cite{niu2018hybrid}. 

Prior data placement techniques~\cite{matsui2017design,sun2013high,heuristics_hyrbid_hystor_sc_2011,vasilakis2020hybrid2,lv2013probabilistic,li2014greendm,guerra2011cost,elnably2012efficient,heuristics_usenix_2014,doudali2019kleio,ren2019archivist,cheng2019optimizing,raghavan2014tiera,salkhordeh2015operating,hui2012hash,xue2014storage,zhang2010automated,zhao2010fdtm,shi2013optimal,wu2012data,ma2014providing,iliadis2015exaplan,wu2009managing,wu2010exploiting,park2011hot} propose hand-crafted policies that (i) consider only a limited number of workload characteristics (e.g., access frequency or access recency) to identify the storage device for an incoming I/O request~\cite{matsui2017design,sun2013high,heuristics_hyrbid_hystor_sc_2011,vasilakis2020hybrid2,lv2013probabilistic,li2014greendm,guerra2011cost,elnably2012efficient,heuristics_usenix_2014,lv2013hotness,montgomery2014extent}, and (ii) do not consider the variations in read/write latencies of storage devices~\cite{matsui2017design,sun2013high,heuristics_hyrbid_hystor_sc_2011,vasilakis2020hybrid2,lv2013probabilistic,li2014greendm,guerra2011cost,elnably2012efficient,heuristics_usenix_2014,doudali2019kleio,ren2019archivist}, or the number and types of storage devices in the HSS~\cite{matsui2017design,sun2013high,heuristics_hyrbid_hystor_sc_2011,lv2013probabilistic,li2014greendm,guerra2011cost,elnably2012efficient,heuristics_usenix_2014,ren2019archivist, matsui2017tri, matsui2017design}. 
As a result, such data placement techniques cannot easily adapt to the dynamic workload demands and diverse real-world hybrid storage system configurations. 
To demonstrate this lack of adaptivity, we show the average request latencies of two prior data-placement techniques, \chg[1]{CDE~\cite{matsui2017design} and RNN-HSS~\cite{doudali2019kleio},} for four representative workloads on a performance-oriented HSS in~\Cref{fig:motivation}(e) and cost-oriented HSS in~\Cref{fig:motivation}(f). 
We add three boundary scenarios to our evaluation: (1) Slow-Only, where all the data is placed in the slow storage device, (2) Fast-Only, where all the data is placed in the fast storage device, and (3) Oracle~\cite{meswani2015heterogeneous}, which performs data placement with complete knowledge of future I/O access patterns.

We make two key observations. 
First, CDE and RNN-HSS show a large average performance loss compared to Oracle: 41.1\% (32.6\%) and 34.4\% (47.6\%) in performance-oriented (cost-oriented) HSS, respectively. 
While both techniques achieve comparable performance to Oracle in select workloads (e.g., \texttt{usr\_0} for CDE in cost-oriented HSS, \texttt{hm\_1} for RNN-HSS in performance-oriented HSS), they show sub-optimal performance in other workloads, showing limited adaptability to workload diversity.
Second, both techniques demonstrate inconsistent behavior across the two HSS configurations: in \texttt{hm\_1} workload, \chg[1]{\rn{both CDE and RNN-HSS}} underperform the Slow-Only in performance-oriented HSS, but outperform \chg[1]{the Slow-Only} in cost-oriented HSS. This shows the inability of prior approaches to adapt to varying storage device characteristics and system heterogeneity.  

We conclude that prior hand-crafted data placement techniques fail to adapt to diverse workload demands or changes in storage device characteristics due to their rigid design choices and limited system awareness.

\section{A PRIMER ON REINFORCEMENT AND PERCEPTRON LEARNING}

\subsection{What is Reinforcement Learning?}

\begin{revblock}
Reinforcement learning (RL)~\cite{rl_bible} is the algorithmic approach to learn how to take an \emph{action} in a given \emph{situation} to maximize a numerical \emph{reward}.
A typical RL system consists of two main components: the agent and
the environment, as illustrated in Fig.~\ref{fig:rl_basics}. 
At every timestep $t$, the agent observes the current \emph{\textbf{state}} of the environment $S_t$ and takes an \emph{\textbf{action}} $A_t$, for which the environment provides a \emph{\textbf{reward}} $R_{t+1}$.
The agent’s goal is to find the optimal \emph{\textbf{policy}} that maximizes the cumulative reward collected from the environment over time.
The expected cumulative reward by taking an action $A$ in a given state $S$ is defined as the \emph{\textbf{Q-value}} of the state-action pair (denoted as $Q(S,A)$).
The agent iteratively optimizes its policy in two steps: (1) updating Q-value using the reward collected from the environment, and (2) optimizing the policy using the updated Q-value.

\begin{figure}[!ht]
    \centering
    \includegraphics[scale=0.4]{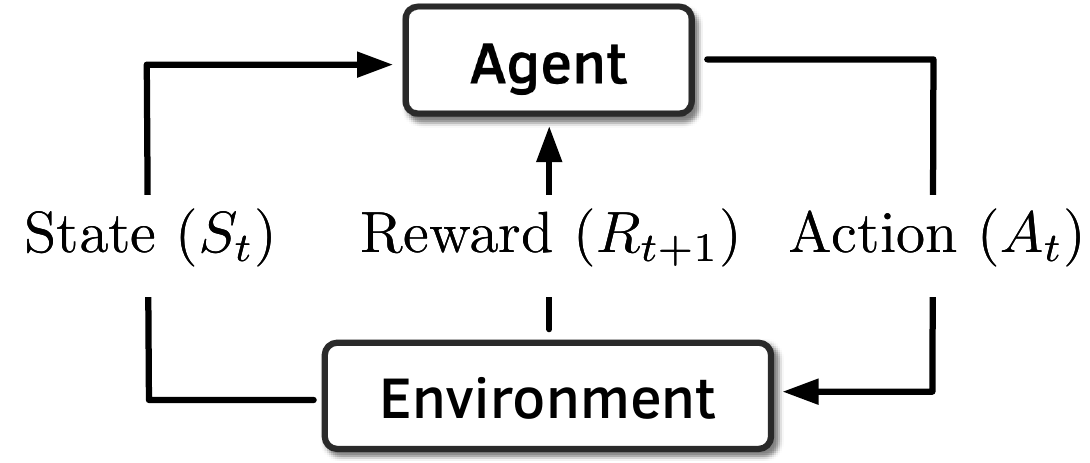}
    \caption{Overview of a reinforcement learning system.}
    \label{fig:rl_basics}
\end{figure}

\paraheading{Updating Q-Values.} Prior research has proposed numerous algorithms to update the Q-values with varying computational complexities~\cite{rl_bible}. 
SARSA is one such algorithm that strikes a good trade-off between the learning accuracy and the computational complexity, which makes it suitable for \emph{online} architectural decision making.
In SARSA, if at a given timestep $t$, the agent observes a state $S_t$, takes an action $A_t$, while the environment transitions to a new state $S_{t+1}$ and emits a reward $R_{t+1}$ and the agent takes action $A_{t+1}$ in the new state, the Q-value of the old state-action pair $Q(S_t, A_t)$ is iteratively optimized using Eq.~\eqref{eq:sarsa}:

\begin{equation}\label{eq:sarsa}
\begin{aligned}
Q\left(S_t, A_t\right) & \gets Q\left(S_t, A_t\right)\\
&+ \alpha\left[R_{t+1}+\gamma Q\left(S_{t+1}, A_{t+1}\right) - Q\left(S_t, A_t\right)\right]
\end{aligned}
\end{equation} 

Here, $\alpha$ is the \emph{learning rate} parameter \rbc{that} controls the convergence rate of Q-values, and $\gamma$ is the \emph{discount factor} that controls the ``far-sighted" planning capability of the agent.

\paraheading{Optimizing Policy.} To maximize cumulative rewards over time, a purely greedy agent always exploits the action providing the highest Q-value. 
However, greedy exploitation may leave the state-action space under-explored. 
Thus, to balance exploration and exploitation, an $\epsilon$-greedy agent randomly selects an action with a small probability 
$\epsilon$ (exploration rate); otherwise, it chooses the action with the highest Q-value.

Prior works have applied RL for various microarchitectural decision-making processes, including main memory scheduling~\cite{rlmc,morse}, cache replacement~\cite{imitation,chrome,rl_cache} and software-hinted hardware prefetching~\cite{peled_rl}. In this article, we describe Pythia, which is the first RL-based software-transparent hardware prefetching mechanism.

\subsection{What is Perceptron Learning?}

Perceptron learning is a simplified learning model to mimic biological neurons.
Fig.~\ref{fig:perc_basic} shows a \emph{single-layer perceptron} network where each \emph{input} is connected to the \emph{output} via an \emph{artificial neuron}. Each artificial neuron is represented by a numeric value, called \emph{weight}. 
The perceptron network as a whole iteratively learns a binary classification function $f(x)$ (shown in Eq.~\ref{eq:perceptron}), a function that maps the input $X$ (a vector of $n$ values) to a single binary output.

\begin{figure}[!ht]
    \centering
    \includegraphics[width=\columnwidth]{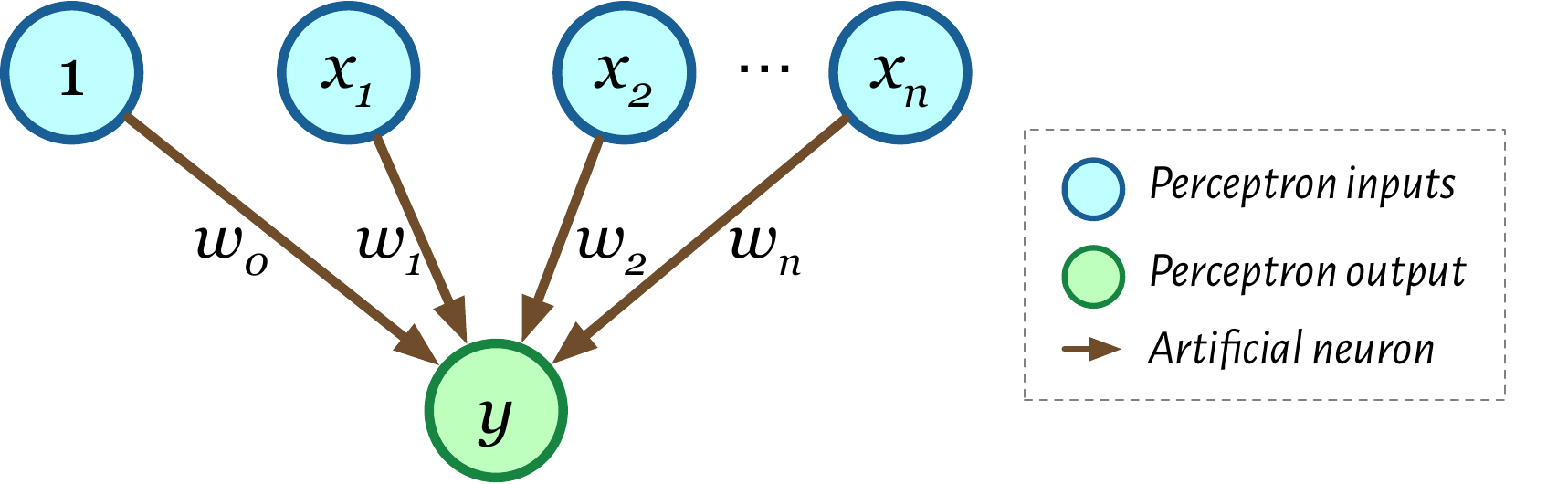}
    \caption{Overview of a single-layer perceptron model.}
    \label{fig:perc_basic}
\end{figure}

\begin{equation} \label{eq:perceptron}
f(x)=\left\{\begin{array}{ll}
1 & \text { if } w_{0}+\sum_{i=1}^{n} w_{i} x_{i}>0 \\
0 & \text { otherwise }
\end{array}\right.
\end{equation}

The perceptron learning algorithm starts by initializing the weight of each neuron and iteratively trains the weights using each input vector from the training dataset in two steps.
First, for an input vector $X$, the perceptron network computes a binary output using Eq.~\ref{eq:perceptron} and the current weight values of its neurons. Second, if the computed output differs from the desired output for that input vector provided by the dataset, the weight of each neuron is updated. This iterative process is repeated until the error between the computed and desired output falls below a user-specified threshold.

Prior research has successfully demonstrated perceptron learning for predicting branch direction~\cite{perceptron,jimenez2016multiperspective,tarjan2005merging,jimenez2003fast,jimenez2002neural}, branch target~\cite{garza2019bit}, branch confidence~\cite{akkary2004perceptron}, cache reuse~\cite{jimenez2017multiperspective,teran2016perceptron}, prefetch usefulness~\cite{ppf,jamet2024tlp}. In this article, we \chg[1]{describe} Hermes, which is the first work that applies perceptron learning for off-chip prediction.

\end{revblock}

\section{DATA PREFETCHING USING REINFORCEMENT LEARNING}

\subsection{Formulating Prefetching as an RL Agent}

\begin{figure*}[!ht]
    \centering
    \includegraphics[width=0.9\textwidth]{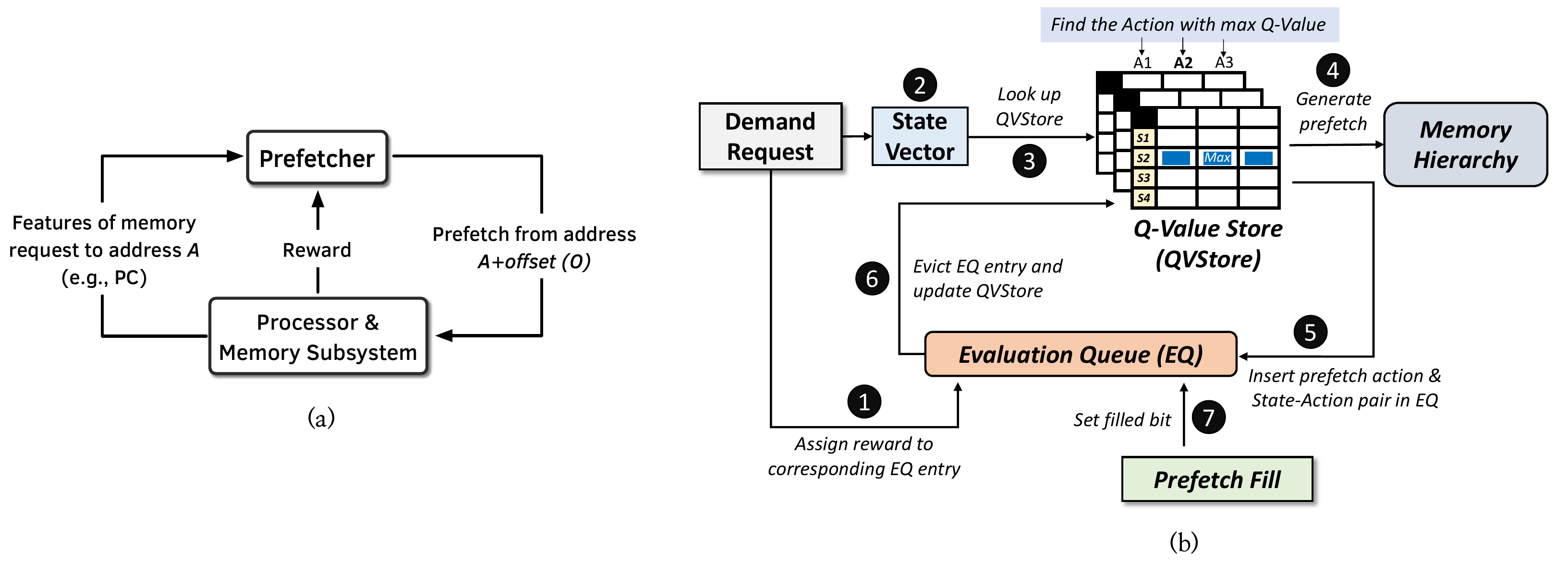}
    \caption{(a) Formulating prefetcher as an RL agent. (b) Overview of Pythia. \chg[1]{Figures adapted from our MICRO 2021 paper~\cite{pythia}.}}
    \label{fig:pythia_main}
\end{figure*}

We formulate prefetching as a reinforcement learning (RL) problem, as shown in~\Cref{fig:pythia_main}(a). 
Specifically, Pythia is an RL-agent that learns accurate, timely, and system-aware prefetch decisions by interacting with the processor and memory subsystem. 
At each timestep, corresponding to a new demand request, Pythia observes system state and takes a prefetch action. 
Each action (including \chg[1]{not issuing a prefetch}) yields a numerical reward based on the accuracy and timeliness of the prefetch request, and various system-level feedback \chg[1]{information}. 
Pythia's goal is to identify an optimal policy maximizing accurate and timely prefetches, while taking system-level
feedback information into account. While Pythia's framework can be extended to various types of system-level feedback, \chg[1]{in this work} we demonstrate Pythia using memory bandwidth usage as a feedback.

\paraheading{State.} We define the state as a $k$-dimensional vector of program features, each comprising \chg[1]{of} two components: (1) control-flow, and (2) data-flow. 
The control-flow component includes basic information like load-PC or branch-PC, and a history indicating whether it is derived from the current or past requests. 
The data-flow component includes cacheline address, physical page number, page offset, cacheline delta, and associated history. 
Although Pythia can utilize many features, we fix the state-vector size (i.e., $k$) at design time given a limited hardware storage budget. 
However, the exact set of $k$ features is configurable online via configuration registers.

\begin{revblock}
\paraheading{Action.} We define the action of the RL-agent as selecting a prefetch offset (i.e., a delta between the predicted and the demanded cacheline address) from a set of candidate prefetch offsets. 
To limit prefetch requests within the physical page of the triggering demand request, the list of prefetch offsets only contains values in the range of [-63,63] for a system with a traditionally-sized $4$KB page and $64$B cacheline. A zero offset is also a valid action for Pythia that signifies no prefetch request is generated. 
\end{revblock}

\paraheading{Reward.} 
The reward structure defines the prefetcher's objective. 
We define five different reward levels:
\begin{itemize}
    \item \textbf{\emph{Accurate and timely}} ($\mathcal{R}_{AT}$) reward, assigned to an action whose corresponding prefetch address gets demanded \emph{after} the prefetch fill.
    \item \textbf{\emph{Accurate but \rbc{late}}} ($\mathcal{R}_{AL}$) reward, assigned to an action whose corresponding prefetch address gets demanded \emph{before} the prefetch fill.
    \item \textbf{\emph{Loss of coverage}} ($\mathcal{R}_{CL}$) reward, assigned to an action whose corresponding prefetch address \rbc{is to a different physical page than the triggering demand request.}
    \item \textbf{\emph{Inaccurate}} ($\mathcal{R}_{IN}$) reward, assigned to an action whose corresponding prefetch address does \emph{not} get demanded in a temporal window. The reward is classified into two sub-levels: inaccurate given low bandwidth \rbc{usage} ($\mathcal{R}_{IN}^{L}$) and inaccurate given high bandwidth \rbc{usage} ($\mathcal{R}_{IN}^{H}$).
    \item \textbf{\emph{No-prefetch}} ($\mathcal{R}_{NP}$) reward, assigned when Pythia decides not to prefetch. This reward level is also classified into two sub-levels: no-prefetch given low bandwidth \rbc{usage} ($\mathcal{R}_{NP}^{L}$) and no-prefetch given high bandwidth \rbc{usage} ($\mathcal{R}_{NP}^{H}$).
\end{itemize}

\begin{revblock}
\chg[1]{The reward levels, in unison, provide prefetching objective to Pythia}.
$\mathcal{R}_{AT}$ and $\mathcal{R}_{AL}$ guide Pythia toward accurate and timely prefetches. 
$\mathcal{R}_{CL}$ encourages prefetching within the same physical page as the triggering request. 
$\mathcal{R}_{IN}$ and $\mathcal{R}_{NP}$ shape Pythia’s strategy based on memory bandwidth usage. 
\end{revblock}

\subsection{Pythia: Design Overview}

\Cref{fig:pythia_main}(b) shows a high-level overview of Pythia, comprising two primary hardware structures: \emph{Q-Value Store} (QVStore) and \emph{Evaluation Queue} (EQ). QVStore records Q-values for observed state-action pairs. EQ maintains a FIFO list of recently taken actions, where each entry contains three information: (1) the selected action, (2) corresponding prefetch address, and (3) a \emph{filled} bit indicating cache fill status.

Upon receiving a demand request, Pythia checks the EQ for the requested memory address (\circled{1}). 
If present, it implies a previously issued prefetch was useful, prompting Pythia to assign a reward (either $\mathcal{R}_{AT}$ or $\mathcal{R}_{AL}$), depending on the filled bit. 
Next, Pythia constructs a state-vector from the demand request attributes (e.g., PC, address, cacheline delta) (\circled{2}) and consults QVStore to identify the action with the highest Q-value (\circled{3}). 
Pythia selects this action and issues the corresponding prefetch request to the cache hierarchy. 
Simultaneously, Pythia inserts the chosen action, its prefetch address, and state-vector into EQ (\circled{5}). 
Actions involving no prefetch or addresses outside the current physical page are also inserted, with rewards assigned immediately. 
Upon eviction of an EQ entry, Pythia updates Q-values using the stored state-action pair and reward (\circled{6}). 
When a prefetch request gets filled in cache, Pythia sets the filled bit in its EQ entry (\circled{7}), marking it as timely or late.

\subsubsection{\textbf{\edit{Storage Overhead}}} 
\edit{Overall, Pythia requires only $25.5$~KB of storage \chg[1]{per core}, in which QVStore and EQ consume $24$~KB and $1.5$~KB, respectively}.

\subsection{Evaluation Methodology}

We use the ChampSim~\cite{champsim} trace-driven simulator to evaluate Pythia.
We simulate an Intel Skylake~\cite{skylake}-like multi-core processor that supports up to $12$ cores.
For single-core (multi-core) simulations, we warm up the core using $100$ million ($50$ million) instructions, followed by simulating the next $500$ million ($150$ million) instructions. 
We evaluate Pythia using a diverse set of $150$ memory-intensive workload traces spanning \texttt{SPEC CPU2006}~\cite{spec2006}, \texttt{SPEC CPU2017}~\cite{spec2017}, \texttt{PARSEC}~\cite{parsec}, \texttt{Ligra}~\cite{ligra}, and \texttt{Cloudsuite}~\cite{cloudsuite} benchmark suites, and compare it against five state-of-the-art prior prefetchers: SPP~\cite{spp}, SPP+PPF~\cite{ppf}, SPP+DSPatch~\cite{dspatch}, Bingo~\cite{bingo}, and MLOP~\cite{mlop}.
All prefetchers are trained on L1-cache misses and fill
prefetched lines into L2 and last-level cache (LLC).
\edit{We also prototype Pythia using Chisel HDL~\cite{chisel} to accurately estimate its area, power, and latency overhead, and incorporate them in the performance model}.
We open-source our artifact-evaluated implementation of Pythia, and all necessary evaluation infrastructure to facilitate future research~\cite{pythia_github}.

\subsection{Key Results}

\subsubsection{\textbf{Performance Evaluation \chg[1]{with} Varying Number of Cores}}
\begin{figure}[!ht]
    \centerline{\includegraphics[width=\columnwidth]{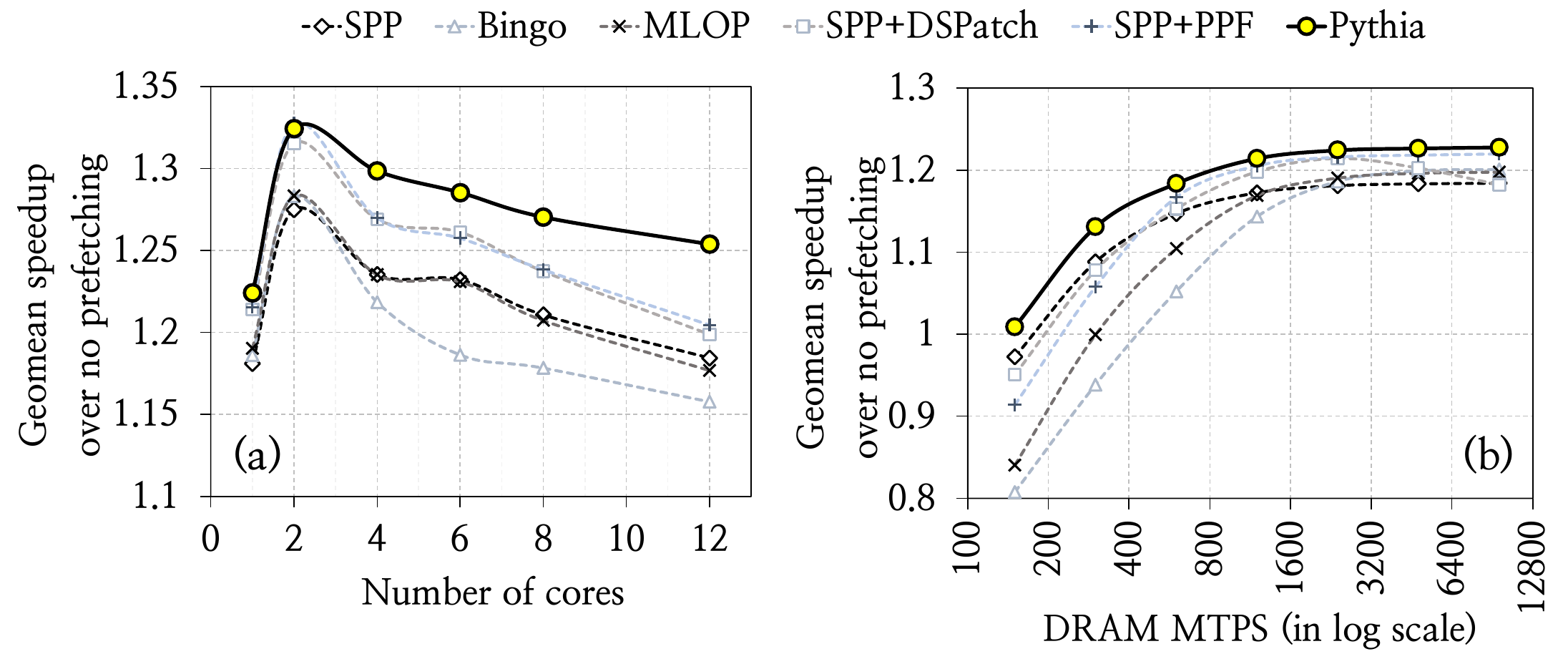}}
    \caption{Average performance improvement of \chg[1]{prior prefetchers and Pythia} in systems with varying (a) number of cores and (b) DRAM million
transfers per second (MTPS). \chg[1]{Figures adapted from our MICRO 2021 paper~\cite{pythia}}.}
    \label{fig:pythia_eval}
\end{figure}

\Cref{fig:pythia_eval}(a) illustrates the average performance improvement across all traces for prefetchers in single-core to 12-core systems. 
We make two key observations. 
First, Pythia consistently outperforms MLOP, Bingo, and SPP across \emph{all} configurations. 
Second, Pythia's performance gain over \chg[1]{all evaluated} prior prefetchers increases with an increase in the number of processor cores.
\edit{As core count increases, per-core memory bandwidth becomes increasingly constrained, limiting the effectiveness of bandwidth-unaware prefetchers. By explicitly adapting its prefetching policy based on bandwidth usage, Pythia achieves larger performance gains as core count increases.}
In single-core systems, Pythia outperforms MLOP, Bingo, SPP, and aggressive SPP with perceptron filtering (PPF) by $3.4$\%, $3.8$\%, $4.3$\%, and $1.02$\%, respectively. For four-core (and twelve-core) systems, Pythia exceeds MLOP, Bingo, SPP, and SPP+PPF by $5.8$\% ($7.7$\%), $8.2$\% ($9.6$\%), $6.5$\% ($6.9$\%), and $3.1$\% ($5.2$\%), respectively.

\subsubsection{\textbf{Performance Evaluation \chg[1]{with} Varying DRAM Bandwidth}}
To evaluate Pythia for bandwidth-constrained commercial server-class processors (where each core receives only a fraction of a DRAM channel), we simulate a single-core, single-channel configuration by scaling DRAM bandwidth (\Cref{fig:pythia_eval}(b)). 
Pythia consistently outperforms all competing prefetchers across bandwidth levels ranging from $\frac{1}{16}\times$ to $4\times$ the baseline. MLOP and Bingo suffer substantial performance drops under low \chg[1]{DRAM} bandwidth due to overprediction. By trading prefetch coverage for accuracy based on bandwidth usage, Pythia outperforms MLOP, Bingo, SPP, and SPP+PPF by $16.9$\%, $20.2$\%, $3.7$\%, and $9.5$\%, respectively, in the most constrained $150$-MTPS configuration. Even under ample bandwidth at $9600$-MTPS, Pythia leads MLOP, Bingo, SPP, and SPP+PPF by $3$\%, $2.7$\%, $4.4$\%, and $0.8$\%, respectively.

These results succinctly demonstrate that Pythia's RL-based prefetching policy dynamically adapts to a wide range of workloads and system configurations, providing \emph{consistent} performance gains over many prior-best human-designed prefetchers.
\chg[1]{Additional results can be found in our MICRO 2021 paper~\cite{pythia}.}

\section{OFF-CHIP PREDICTION VIA PERCEPTRON LEARNING}

\subsection{Hermes: Design Overview}

\begin{figure*}[!ht]
    \centering
    \includegraphics[width=0.9\textwidth]{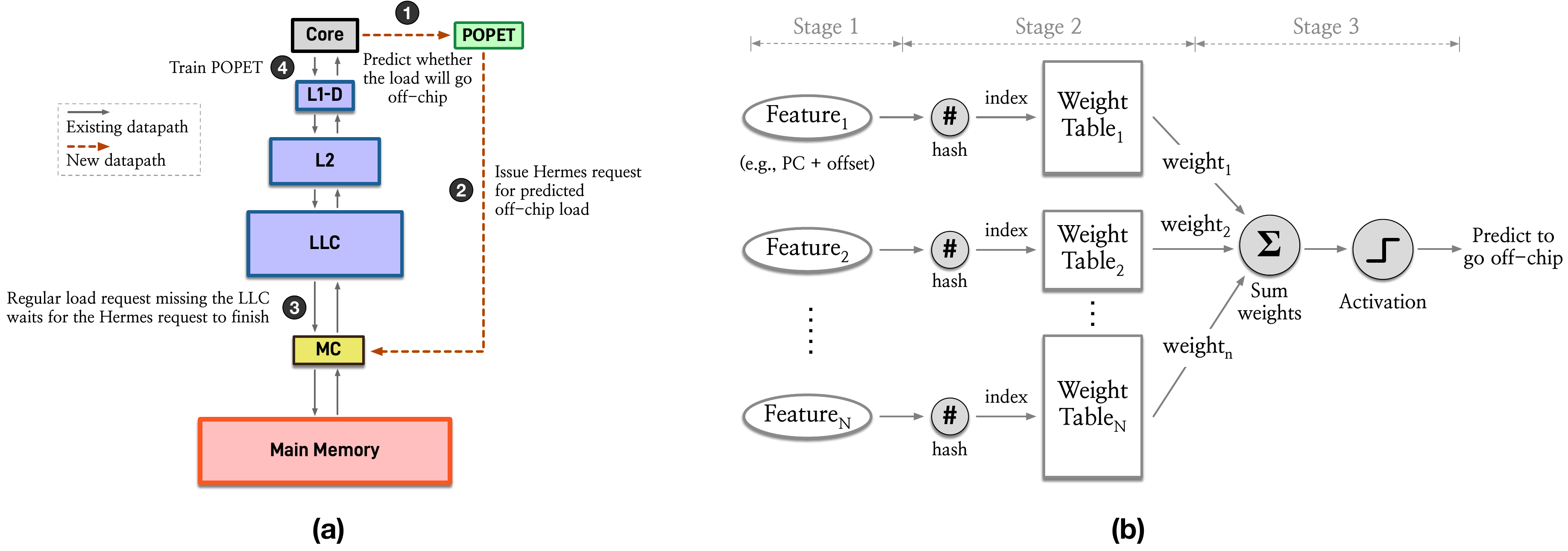}
    \caption{(a) Overview of Hermes. (b) \chg[1]{Hardware} stages to make a prediction by POPET. \chg[1]{Figures adapted from our MICRO 2022 paper~\cite{hermes}.}}
    \label{fig:hermes_overview}
\end{figure*}

\Cref{fig:hermes_overview}(a) provides a high-level overview of Hermes~\cite{hermes}.
POPET is the key component of Hermes that accurately predicts \chg[1]{which} load requests \chg[1]{will go off-chip (i.e., access DRAM)} by correlating a load request with multiple program features via perceptron learning.
For each processor-generated demand load \chg[1]{instruction}, POPET predicts whether the request will go off-chip (\circled{1}). 
If so, Hermes issues a speculative memory request (called a \emph{Hermes request}) \emph{directly} to the main memory controller immediately after obtaining the physical address (\circled{2}). 
This Hermes request proceeds in parallel with the regular load accessing the on-chip cache. 
If the prediction is correct, the regular load eventually misses in the LLC and waits for the ongoing Hermes request, effectively hiding on-chip cache access latency (\circled{3}). 
If the Hermes request completes but no regular load targets the same address \chg[1]{(e.g., a predicted off-chip load request hitting on-chip cache)}, Hermes discards it, avoiding unnecessary cache fills and preserving coherence. 
Hermes uses each \chg[1]{retired} load to train POPET for future predictions (\circled{4}).

\subsection{Design of \chg[1]{the POPET Off-Chip Load Predictor}}

\begin{revblock}
POPET is composed of multiple one-dimensional weight tables, each associated with a single program feature. 
Each table entry holds a $5$-bit saturating signed integer that \chg[1]{records} the correlation between a feature value and the actual outcome (i.e., whether the load actually went off-chip). 
A weight value closer to $+15$ or $-16$ indicates strong positive or negative correlation, respectively, while values near zero indicate weak correlation. 
During \chg[1]{the} training (step \circled{4} in~\Cref{fig:hermes_overview}(a)), these weights are updated based on the true outcome to refine POPET’s predictions. 
\end{revblock}

\subsubsection{\textbf{Making Predictions using POPET}}
During load queue (LQ) allocation for a core-generated load (step \circled{1} in~\Cref{fig:hermes_overview}(a)), POPET performs a binary prediction on whether the load will go off-chip, following a three-stage process shown in~\Cref{fig:hermes_overview}(b). 
First, POPET extracts various program features from the current load and prior requests. 
Second, it hashes each feature value to index into the corresponding weight table and retrieves the weight. 
Third, it sums all weights to compute the cumulative perceptron weight ($W_{\sigma}$). 
If $W_{\sigma}$ exceeds the activation threshold ($\tau_{act}$), POPET predicts that the load will go off-chip. 
The hashed feature values, $W_{\sigma}$, and the prediction are stored in the LQ entry to train POPET when the load completes (step \circled{4} in~\Cref{fig:hermes_overview}(a)).

\subsubsection{\textbf{Training POPET}}
POPET training begins when a demand load returns to the core and releases its load queue (LQ) entry (step \circled{4} in~\Cref{fig:hermes_overview}(a)). Loads that miss the LLC and access main memory are labeled as true off-chip requests. POPET uses this true outcome and the stored prediction in the LQ entry to update feature weights in two stages. First, the previously computed $W_{\sigma}$ is retrieved. If $W_{\sigma}$ lies between the positive and negative training thresholds, $T_P$ and $T_N$, respectively, training is triggered. 
This check prevents over-saturation of weights and enables quick adaptation to program phase changes. 
Second, if training proceeds, POPET retrieves each feature’s weight using the hashed indices from the LQ entry. 
If the true outcome is positive (off-chip), the weights are incremented by one; if negative, they are decremented by one. 
This update mechanism steers each feature weight toward the true outcome, improving prediction accuracy over time.

\subsubsection{\textbf{\edit{Storage Overhead}}}
\edit{Overall, Hermes requires only $4$~KB of storage \chg[1]{per core}, in which POPET and LQ metadata consume $3.2$~KB and $0.8$~KB, respectively~\cite{hermes}}.

\subsection{Evaluation Methodology}

We use the ChampSim~\cite{champsim} trace-driven simulator to evaluate Hermes. 
We faithfully model the latest-generation Intel Alder
Lake performance-core with its large \chg[1]{reorder buffer}, large caches
with publicly-reported on-chip cache access latencies.
For single-core (multi-core) simulations, we warm up the core using $100$ million ($50$ million) instructions, followed by simulating the next $500$ million ($100$ million) instructions. 
We evaluate Hermes using $110$ memory-intensive workload traces spanning \texttt{SPEC CPU2006}~\cite{spec2006}, \texttt{SPEC CPU2017}~\cite{spec2017}, \texttt{PARSEC}~\cite{parsec}, \texttt{Ligra}~\cite{ligra},
and commercial workloads from the 2nd data value prediction championship (\texttt{CVP}~\cite{cvp2}),
and compare Hermes against (1) five state-of-the-art data prefetchers (i.e., Pythia~\cite{pythia}, Bingo~\cite{bingo}, SPP with PPF~\cite{spp,ppf}, MLOP~\cite{mlop}, and SMS~\cite{sms}), and (2) two prior off-chip prediction mechanisms: hit/miss predictor (HMP~\cite{yoaz1999speculation}) and cacheline tag-tracking predictor (TTP~\cite{lp,hermes}).
\edit{We also evaluate Hermes using a range of latencies for issuing a Hermes request directly to the memory controller (step \circled{2} in~\Cref{fig:hermes_overview}(a)) to account for diverse on-chip interconnect designs}.
We open-source our artifact-evaluated implementation of Hermes, and all necessary evaluation infrastructure to facilitate future research~\cite{hermes_github}.

\subsection{Key Results}

\subsubsection{\textbf{Performance Evaluation \chg[1]{with Different} Off-Chip Predictors}}
\begin{figure}[!ht]
    \centering
    \includegraphics[width=\columnwidth]{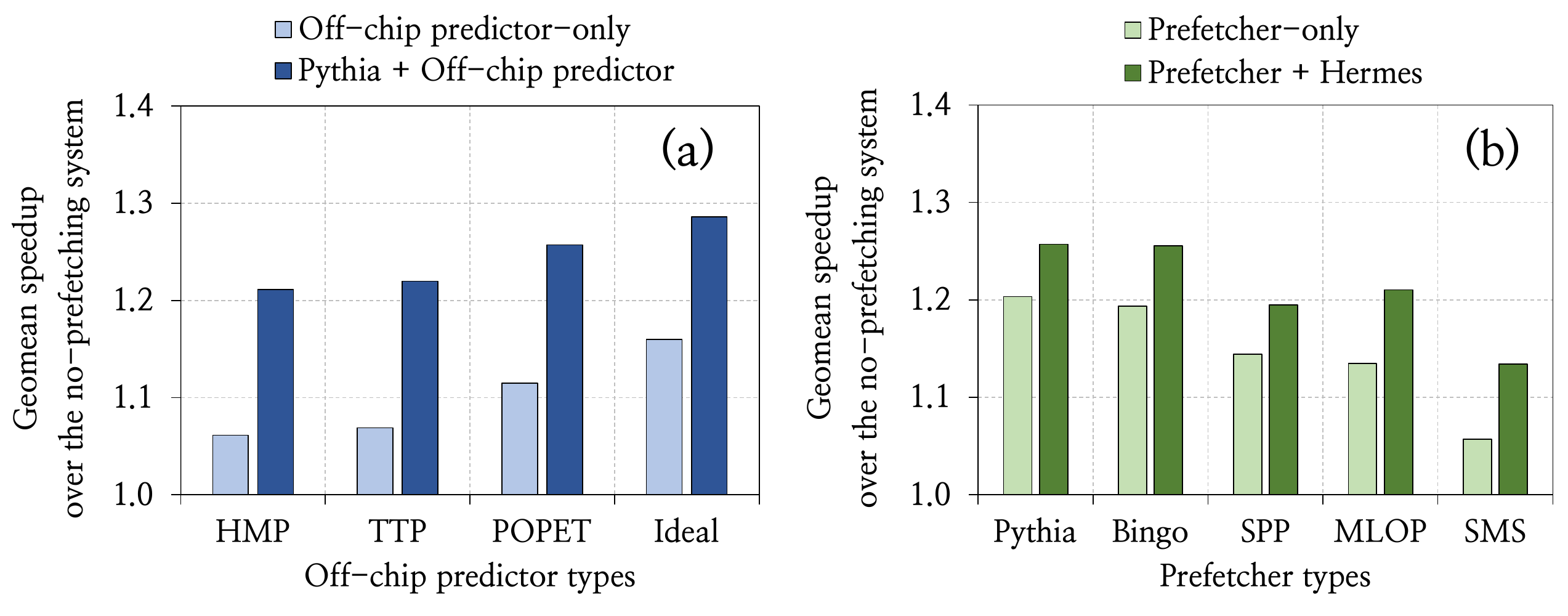}
    \caption{Performance comparison by varying (a) off-chip predictors and (b) the underlying prefetcher. \chg[1]{Figures adapted from our MICRO 2022 paper~\cite{hermes}.}}
    \label{fig:hermes_eval}
\end{figure}

\Cref{fig:hermes_eval}(a) shows the performance of Hermes with POPET, HMP, TTP, and an ideal off-chip predictor, with and without an underlying baseline prefetcher (in this case, Pythia) and normalized to a no-prefetching system for single-core workloads. 
We highlight two key observations. 
First, Hermes with POPET outperforms Hermes-HMP and Hermes-TTP. \edit{This is due to POPET's high accuracy off-chip prediction ($77.1\%$ on average) as compared to HMP ($47\%$) and TTP (\chg[1]{$16.6\%$)}}.\footnote{\edit{Our MICRO 2022 paper~\cite{hermes} further shows that POPET’s high-accuracy off-chip prediction incurs minimal overhead: increasing main memory requests by $5.5\%$ and dynamic power by $3.6\%$ over no prefetching}.}
On average, Hermes-HMP, Hermes-TTP, and Hermes with POPET improve performance over Pythia by $0.8$\%, $1.7$\%, and $5.4$\%, respectively. 
Second, Hermes-POPET achieves nearly $90$\% of the performance gain of the ideal off-chip predictor that assumes perfect prediction. 
These results show that POPET's high prediction accuracy and coverage are central to Hermes's effectiveness, underscoring the importance of a well-designed off-chip predictor.

\subsubsection{\textbf{Performance Evaluation \chg[1]{with Different} Underlying Prefetchers}}
We evaluate Hermes \chg[1]{naively}-combined (\chg[1]{i.e., without any implicit coordination}) with five state-of-the-art data prefetchers at last-level cache. \Cref{fig:hermes_eval}(b) shows the performance normalized to a no-prefetching single-core baseline. Hermes combined with any prefetcher \emph{consistently} outperforms the prefetcher alone. Specifically, Hermes improves performance over Bingo, SPP, MLOP, and SMS by $6.2$\%, $5.1$\%, $7.6$\%, and $7.7$\%, respectively.
\chg[1]{Our recent work, Athena~\cite{athena}, shows that an RL-based coordination policy to synergize off-chip predictor with prefetcher can provide even more performance improvement than their naive combination.}

These results suggest that by employing a learning-based, adaptive off-chip prediction mechanism, Hermes \emph{consistently} outperforms prior off-chip predictors across diverse workloads and system configurations.
\chg[1]{More results can be found in our MICRO 2022 paper~\cite{hermes}.}

\section{DATA PLACEMENT IN HYBRID STORAGE SYSTEMS USING REINFORCEMENT LEARNING}

\begin{revblock}
We propose Sibyl~\cite{singh2022sibyl}, a reinforcement learning (RL)-based data placement technique for hybrid storage systems.
While RL is a promising alternative to existing data placement techniques, its effectiveness depends on how the data placement problem is cast as an RL task.

\subsection{Formulating Data Placement as an RL Problem}
\Cref{fig:sibyl_rlformulation_design}(a) shows our formulation of  data placement as an RL problem. We design Sibyl as an RL agent that learns to perform accurate and system-aware data placement decisions by interacting with the hybrid storage system. 
For each storage request, Sibyl observes multiple workload and system-level features as a \emph{state} to make a placement decision. After every \emph{action}, Sibyl receives a \emph{reward} based on the I/O request latency, which reflects both the data placement decision and the HSS state. 
Sibyl aims to learn an optimal data placement policy that improves performance under the current workload and system configuration by minimizing the average request latency. This involves maximizing the use of the fast storage device while avoiding the eviction penalty of non-performance-critical pages. 

\begin{figure*}[!ht]
\centering
\includegraphics[width=0.9\textwidth]{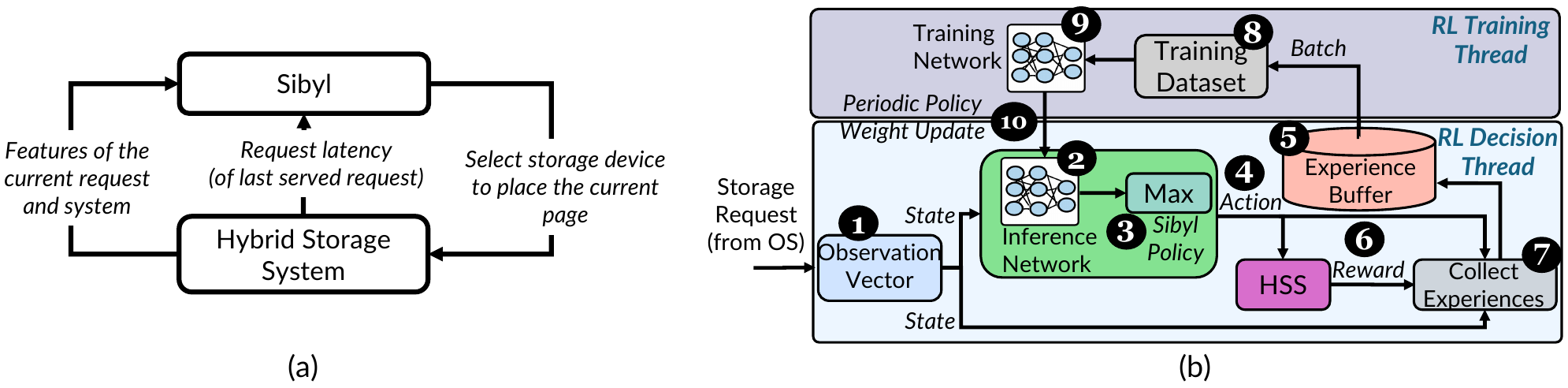}
\caption{(a) Formulation of data placement as an RL Problem. (b) Design of Sibyl. \chg[1]{Figures adapted from our ISCA 2022 paper~\cite{singh2022sibyl}.}}
\label{fig:sibyl_rlformulation_design}
\end{figure*}

\paraheading{State.}
At each time-step $t$, the state features for a particular read/write request are collected in an \emph{observation} vector. 
We perform feature selection to determine the best state features to include in Sibyl's observation vector.
We use a limited number of features to reduce the implementation overhead of our mechanism. Sibyl's  observation vector is a 6-dimensional tuple:
\begin{equation}
O_t = (size_t, type_t, intr_t, cnt_t, cap_t, curr_t).
\end{equation}

\noindent
Table~\ref{tab:state}  
lists our six selected features. We quantize the state features into a small number of bins to reduce the storage overhead.
The current request size ($size_t$) helps distinguish sequential from random access patterns. 
The request type ($type_t$) differentiates between read and write requests, and captures the asymmetry in read and write latencies. 
The access interval ($intr_t$) and access count ($cnt_t$) represent the temporal and spatial locality, respectively. 
$cap_t$ tracks the available capacity in the fast storage, guiding the agent to manage the fast storage capacity while avoiding eviction penalties effectively. 
$curr_t$ is the current placement of the requested page, enabling Sibyl to perform past-aware decisions.

\begin{table}[!ht]
 \caption{State features used by Sibyl}
    \label{tab:state}
\centering
 \renewcommand{\arraystretch}{1}
\setlength{\tabcolsep}{2pt}
  \resizebox{1\linewidth}{!}{%
\begin{tabular}{l||l|c|c}
\hline
\textbf{Feature} & \textbf{Description} & \textbf{\# of bins} & 
\textbf{Encoding (bits)}\\ 
\hline
$size_t$ &  Size of the requested page (in pages) &  8 & 8 \\
$type_t$ &  Type of the current request (read/write)  & 2 & 4 \\
$intr_t$ & Access interval of the requested page & 64 & 8 \\ 
$cnt_t$ & Access count of the requested page & 64 & 8 \\
$cap_t$ &  Remaining capacity in the fast storage device  & 8& 8 \\
$curr_t$ &   Current {placement} of the requested page (fast/slow) & 2 & 4 \\
\hline
\end{tabular}
}
\end{table}

\paraheading{Action.} 
At each time-step $t$, in a given state, Sibyl selects an action from all possible actions. In a hybrid storage system with two devices, possible actions are: placing data in (1) the fast storage device or (2) the slow storage device. This is easily extensible to a large number of storage devices. 

\paraheading{Reward.} 
After every data placement decision at time-step $t$,\footnote{In HSS, a time-step is defined as a new storage request.
} Sibyl gets a reward from the environment at time-step $t+1$, serving as feedback for the previous action.
We craft Sibyl's reward function \textit{R} as follows:
\begin{equation}
R =\begin{cases}
       {{\frac{1}{L_t} }} & 
        \begin{gathered}
            \textit{if no eviction of a page from the} \\[-\jot]
            \textit{fast storage to the slow storage}
        \end{gathered}\\
        max(0,\frac{1}{L_t}-\textit{$R_p$}) & \text{\textit{in case of eviction} }
    \end{cases}
\end{equation}
where $L_t$ and \textit{$R_p$} represent the latency of the last served request and eviction penalty, respectively. If the fast storage is running out of free space, Sibyl evicts non-performance-critical pages to the slow storage. To discourage excessive use of fast storage, we incorporate an eviction penalty (\textit{$R_p$}) into the reward to guide Sibyl to place only performance-critical pages in the fast storage. We empirically set \textit{$R_p$} to 0.001$\times$$L_e$ ($L_e$ is the time required to evict pages from fast to slow storage), which prevents the agent from excessively using the fast storage device.

Sibyl's reward is a function of the request latency because it captures the state of the hybrid storage system, as it significantly varies depending on the request type, device type, and the internal state and characteristics of the device (e.g., read/write latencies, garbage collection latency, queuing delays, and error handling latencies).
\end{revblock}

\subsection{Sibyl: Design Overview}
We implement Sibyl in the storage management layer of the host system's operating system. 
As shown in~\Cref{fig:sibyl_rlformulation_design}(b), Sibyl is composed of two parallel threads: (1) the \emph{RL decision thread}, which performs data placement \circled{4} of the current I/O request while collecting experiences \circled{7} (i.e., actions \circled{4} and rewards \circled{6}) in an \emph{experience buffer} \circled{5}, and 
(2) the \emph{RL training thread}, which uses the collected \textit{experiences}\footnote{Experience is a representation of a transition from one time step to another, in terms of $\langle State, Action, Reward, Next State \rangle$.} \circled{8}  to update its decision-making policy online \circled{9}. 
Sibyl continuously learns from its past decisions and their impact. 
Our two-threaded decoupled implementation ensures that policy training does not interfere with latency-critical placement decisions. To enable parallel execution, we use two separate identical neural networks: an inference network \circled{2} for making placement decisions, and a training network \circled{9} for background learning. Sibyl periodically copies the training network weights to the inference network \circled{10} to adapt the policy for current workload and system conditions. 

For every new storage request to the HSS, Sibyl uses the state information \circled{1} to make a data placement decision \circled{4}. 
The inference network predicts the Q-value for each available action given the state information. Sibyl's policy \circled{3} selects the action with the maximum Q-value and \chg[1]{accordingly} performs the data placement \chg[1]{ (and potentially eviction)}.

\subsection{Evaluation Methodology}
We evaluate Sibyl on a \emph{real system} using two dual HSS configurations: (1) \em{performance-oriented HSS}: high-end device (\textsf{H})~\cite{inteloptane} and middle-end device (\textsf{M})~\cite{intels4510}, and (2) \em{cost-oriented HSS}: high-end device  (\textsf{H})~\cite{inteloptane} and low-end device (\textsf{L})~\cite{seagate}.
\edit{The high-end, middle-end, and low-end devices have a sequential read (write) throughput of $2.4$ ($2$)~GB/s, $550$ ($510$)~MB/s, and $520$ ($450$)~MB/s, respectively.}
The HSS devices appear to the OS as one contiguous logical block address space. 
We implement a lightweight custom block driver to orchestrate I/O requests to the storage devices in HSS. We develop Sibyl using the TF-Agents framework~\cite{TFagents}.
We use fourteen diverse storage traces that have distinct I/O-access patterns (i.e., randomness and hotness properties) from Microsoft Research Cambridge (MSRC)~\cite{MSR}, collected on real enterprise servers.
We compare Sibyl against two heuristic-based (i.e., cold data eviction (CDE~\cite{matsui2017design}) and history-based page selection (HPS~\cite{meswani2015heterogeneous})) and two supervised-learning-based (i.e., Archivist~\cite{ren2019archivist} and RNN-HSS~\cite{doudali2019kleio}) HSS data placement techniques.
We compare the above policies with three boundary scenarios:
(1) Slow-Only, where all data resides in the slow storage, 
(2) Fast-Only, where all data resides in the fast storage, and 
(3) Oracle\rn{~\cite{meswani2015heterogeneous}}, which exploits complete knowledge of future I/O-access patterns to perform optimal data placement.
We open-source Sibyl to facilitate future research~\cite{sibylLink}.

\subsection{Key Results}
\begin{figure}[!ht]
\centering
    \includegraphics[width=\columnwidth]{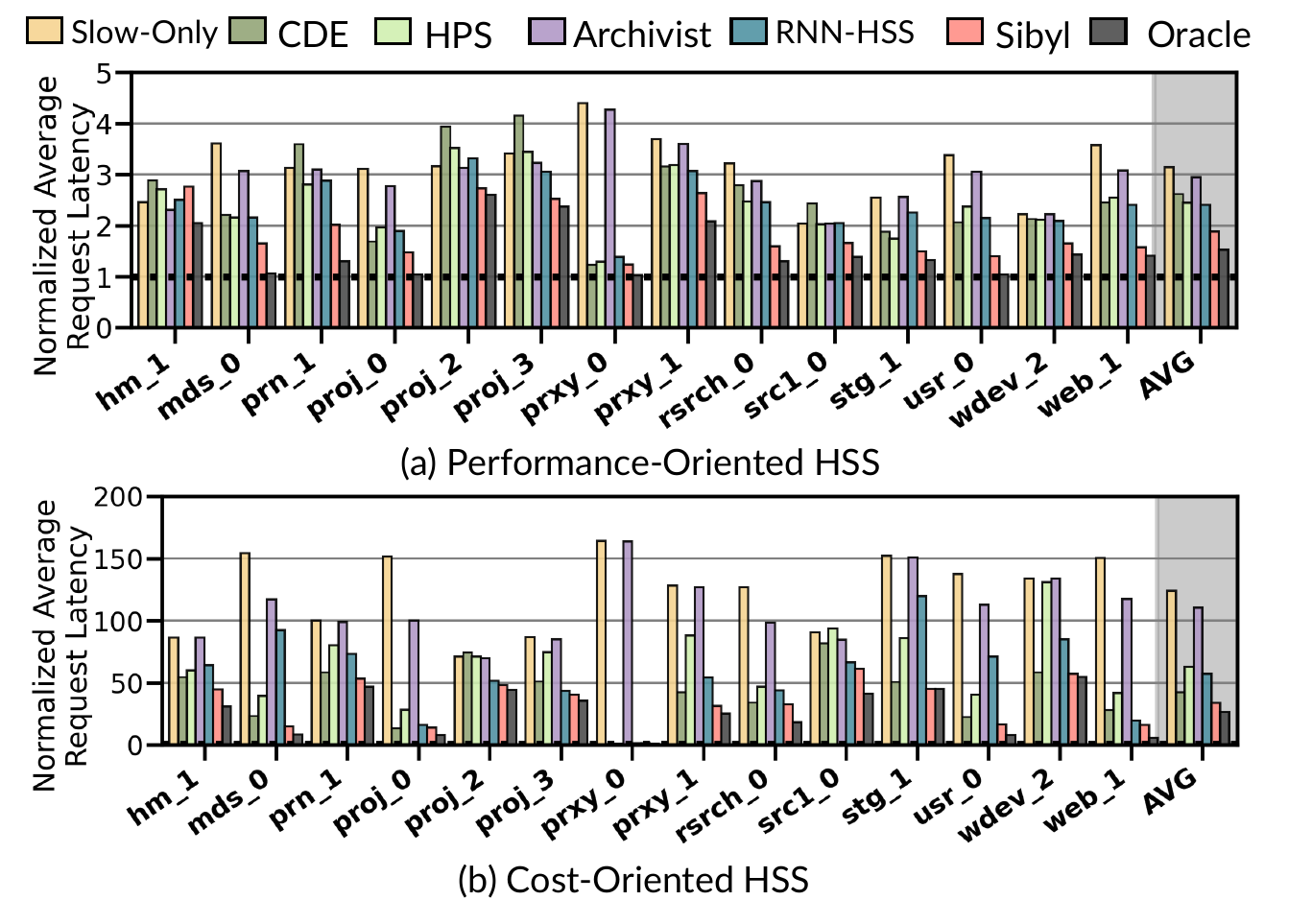}
    \caption{Average request latency under two different hybrid storage configurations (normalized to Fast-Only). \chg[1]{Figures adapted from our ISCA 2022 paper~\cite{singh2022sibyl}.}}
    \label{fig:sibyl_perf}
\end{figure}

\Cref{fig:sibyl_perf} compares the average request latency of Sibyl against the baseline policies for performance-oriented (\Cref{fig:sibyl_perf}(a)) and cost-oriented (\Cref{fig:sibyl_perf}(b)) HSS configurations. All values are normalized to Fast-Only.
We observe that Sibyl consistently outperforms all the baselines for all the \chg[1]{evaluated} workloads in both HSS configurations. 
In the performance-oriented HSS (\Cref{fig:sibyl_perf}(a)), where the latency difference between two devices is relatively smaller than that of cost-oriented HSS, Sibyl improves average performance by 28.1\%, 23.2\%, 36.1\%, and 21.6\% over CDE, HPS, Archivist, and RNN-HSS, respectively. 
In the cost-oriented HSS (\Cref{fig:sibyl_perf}(b)), where there is a large difference between the latencies of the two storage devices, Sibyl improves performance by 19.9\%, 45.9\%, 68.8\%, and 34.1\% over CDE, HPS, Archivist, and RNN-HSS, respectively.
The larger the latency disparity, the greater the benefits of placing only performance-critical pages in the fast storage to avoid evictions.

\chg[1]{We evaluate Sibyl on Tri-HSS with three heterogeneous storage devices. Compared to the state-of-the-art heuristic-based policy, Sibyl improves performance by up to \rn{48.2\%} on average, which demonstrates that Sibyl's adaptive placement policy generalizes to more devices in an HSS. Extending Sibyl to support additional devices requires only minimal changes to the state and action space, highlighting its ease of extensibility.}

We conclude that Sibyl's RL-based data placement technique
consistently outperforms prior \chg[1]{best} data placement techniques across diverse workloads and different HSS configurations. Sibyl's workload- and system-aware online learning policy enables it to adapt dynamically to different conditions, making it a generalizable solution for hybrid storage systems. 
\chg[1]{More results can be found in our ISCA 2022 paper~\cite{singh2022sibyl}.}

\rbone{While Sibyl manages the data placement in HSS, it does
not address the data migration between different storage devices of an HSS. Our recent \rn{proposal, Harmonia~\cite{rakesh2025harmonia}, demonstrates} the first multi-agent RL-based framework for
HSS to manage both data placement and data migration in a synergistic way.}

\section{KEY TAKEAWAYS AND FUTURE OPPORTUNITIES}

This article presents our recent efforts on machine learning (ML)-driven intelligent memory system design.
We present three case studies, Pythia, Hermes, and Sibyl, that employ \chg[1]{reinforcement-learning} and perceptron-learning to enable adaptive, data-driven decision-making across the multi-level cache hierarchies and hybrid storage systems.
Through extensive studies we show that these ML-driven policies consistently outperform prior best human-designed approaches with modest overheads, highlighting the practicality of online learning in real systems.
These case studies also reveal broader opportunities, and challenges, in how such mechanisms are \emph{integrated} and \emph{extended} across the memory hierarchy.

\paraheading{Better Integration \chg[1]{and Coordination} of ML-Driven Mechanisms.}
\edit{Although Pythia, Hermes, and Sibyl improve performance independently and in combination (e.g., Hermes with Pythia), their interactions expose additional opportunities for deeper system-level integration.
We present here two concrete examples of how synergizing these mechanisms can unlock further benefits. 
First, while integrating Hermes with Pythia improves performance on average, we observe that a naive integration often fails to fully realize their combined performance potential.
This motivates designing holistic coordination mechanisms—potentially ML-driven—that jointly coordinate prefetching and off-chip prediction mechanisms.}
\chg[1]{Recently, we have proposed Athena~\cite{athena}, that exploits online reinforcement learning to coordinate off-chip prediction with multiple prefetchers employed throughout the cache hierarchy.}
\rbone{Second, while Sibyl manages data placement within HSS, it does not address the complementary problem of data migration across the heterogeneous storage devices that constitute an HSS. To bridge this gap, our recent proposal, Harmonia~\cite{rakesh2025harmonia}, introduces the first multi-agent RL framework that jointly orchestrates data placement and inter-device data migration, enabling coordinated decision making \rn{in an HSS}. 
By allowing multiple learning agents to collaborate and adapt to dynamic workload characteristics, Harmonia manages these two interdependent tasks synergistically, improving the overall performance of an HSS.}

\paraheading{Extending ML-Driven Design Principles across \chg[1]{the} Memory Hierarchy.}
\edit{The ML-driven design principles presented in this article can be broadly extended for many data-driven decision-making processes across the memory hierarchy, including, but not limited to, (1) co-optimization of caching, prefetching, and memory scheduling mechanisms, \chg[1]{(2) co-optimization of thread and memory scheduling decisions}, (3) data placement and migration in disaggregated memory systems, (4) coordinated data placement and migration in hybrid storage systems.
These problems exhibit complex, non-linear interactions between workload behavior and system state, making them well suited for lightweight, online ML approaches.}

Collectively, these directions highlight the potential of ML-driven approaches to enable adaptive, self-optimizing memory systems that deliver performance and efficiency gains beyond traditional designs,
and we hope this article inspires further research in intelligent \chg[1]{ML-driven} memory system design.

\section*{Acknowledgments}
\rn{We thank all SAFARI Research Group members for providing a stimulating, inclusive, intellectual and scientific environment.
We acknowledge the generous gifts from our funding partners: Futurewei, Google, Huawei, Intel, Microsoft, and VMware.
This work is supported in part by the Semiconductor Research Corporation and the ETH Future Computing Laboratory.}

\bibliographystyle{IEEEtranS}
\bibliography{refs,sibyl_refs}

\end{document}